\documentclass[11pt]{article}
\usepackage{fancyhdr}
\usepackage{url}
\usepackage{times}
\usepackage{graphics}

\newcommand{\eat}[1]{}           

\newcommand{\CKP}{{Check Point FireWall-1}}
\newcommand{\CKPs}{{Check Point firewalls}}
\newcommand{\PIX}{{Cisco PIX}}
\newcommand{\PIXs}{{Cisco PIX firewalls}}
\newcommand{\FC}{{$\cal FC$}}
\newcommand{\RC}{${\cal R}{\cal C}$}

\newtheorem{theorem}{Theorem}[section]
\newtheorem{definition}[theorem]{Definition}

\begin{document}

\newcommand{\mytitle}{Firewall Configuration Errors Revisited}

\title{\mytitle}

\author{Avishai Wool\\
School of Electrical Engineering, \\
Tel Aviv University, Ramat Aviv 69978, ISRAEL \\
E-mail: {\tt yash@acm.org}.
}

\date{\today}

\maketitle



\newcommand{\myabs} { 

  Practically every corporation that is connected to the Internet uses
  firewalls as the first line of its cyber-defense.  However, the
  protection that these firewalls provide is only as good as the
  policy they are configured to implement. The first quantitative
  evaluation of the quality of corporate firewall configurations
  appeared in 2004, based on \CKP\ rule-sets. In general that survey
  indicated that corporate firewalls were often enforcing poorly
  written rule-sets, containing many errors. One important finding
  was that high rule-set complexity was positively correlated with the
  number of detected configuration errors. Another finding was an indication
  that rule-sets from later software versions had slightly fewer
  errors.
 
  The goal of this work is to revisit the first survey, and to test
  whether its findings remain valid. The current study is much
  larger, and is based on newer data, collected from firewalls running
  later firewall versions.  Furthermore, for the first time the study
  includes configurations from two major vendors: both \CKPs\ and
  \PIXs.  Finally, the study considers three times as many possible
  configuration errors, consisting of~36 vendor-neutral errors instead
  of the~12 used in the 2004 study.

  In order to compare the complexity of configurations from different
  vendors, this work also introduces a novel uniform complexity measure,
  called the {\em firewall complexity} (\FC), that applies to both
  types of firewalls.

  The findings of the current study indeed validate the 2004 study's
  main observations: (a) firewalls are (still) poorly configured, and
  (b) a rule-set's complexity, as measured by the new \FC\ measure, is
  (still) positively correlated with the number of detected
  configuration errors.  These findings hold for rule-sets from both
  vendors. Thus we can conclude that, for well-configured firewalls,
  ``small is (still) beautiful''. However, unlike the 2004 study,
  there is no significant indication that later software versions have
  fewer errors (for either vendor). This is apparently because the
  vendor-neutral errors that this study focuses on are all
  controlled by the firewall's basic filtering capability---which has
  not changed significantly between versions.  }

\begin{abstract}
\myabs
\end{abstract}

\pagenumbering {arabic} 

\section{Introduction}

\subsection{Background}

Firewalls are the cornerstones of corporate intranet security.  Once
a firewall is acquired, a firewall administrator has to
configure and manage it to realize an appropriate security policy for
the particular needs of the company.  This is a crucial task; quoting
\cite{rgr97}: ``The single most important factor of your firewall's security
is how you configure it''. 

It is generally accepted among network security experts that
corporate firewalls are poorly configured. Anecdotal evidence of this
sentiment can be found in such mailing lists as the Firewall Wizards
list \cite{fwwiz08}. Furthermore, the poor state of firewall
configuration can be observed, indirectly, by the success of worms and
viruses like Blaster~\cite{CA2003-20} and Saphire~\cite{saph03}, which
could have been easily blocked by a well configured firewall.  This
state of affairs was validated in a 2004 study \cite{woo04-conf},
which for the first time provided a quantitative evaluation of the quality
of corporate firewall configurations collected in 2001.

However, firewall vendors regularly release new 
software versions for their products. In
addition, the attention that is paid to firewall rule-set
quality has increased,
due to regulations such as the Sarbanes-Oxley act \cite{sox}
and the CobiT framework \cite{cobit}; the Payment-Card Industry Data
Security Standard (PCI DSS) \cite{pci}; and the NIST standard 800-41
\cite{nist}.  All these relatively new regulations include specific sections
dealing with firewall configuration, management, and audit.  Thus, one
may hypothesize, or hope, that the quality of corporate firewall
configurations has improved over time.  The goal of this work is to test this
hypothesis, and to check whether the findings of \cite{woo04-conf} are
still valid.

\subsection{Contributions} \label{sec:contrib}

The data for the \cite{woo04-conf} study was collected in 2001, and
clearly much has happened in the network security arena in later years.
Therefore, one may challenge the validity of
\cite{woo04-conf} claiming (or hoping) that the situation has improved
over time. Moreover, \cite{woo04-conf} was fairly small in scope,
covering only~37 rule-sets, all from \CKPs, and considering only~12
possible errors, 8~of which were specific to \CKP.  One could
argue that the sample in \cite{woo04-conf} was not indicative, and
that the detected problems were specific only to that vendor.

To address these possible critiques, the current study has the
following features: (1) It is based on newer configuration data,
collected from firewalls running later firewall versions; (2) It is
significantly larger, covering more than twice as many rule-sets as
\cite{woo04-conf}; (3) For the first time, this study includes
rule-sets from {\em two} leading firewall vendors: \CKP\ (cf.\
\cite{wel02}) and \PIX\ (cf.\ \cite{cf01}); and (4) this study
considers three times as many configuration errors, consisting of~36
vendor-neutral items instead of the~12 used in \cite{woo04-conf}.

The rule-sets were obtained from a variety of corporations 
that used the AlgoSec Firewall Analyzer between 2003--2005
\cite{mwz0x,algosec}.
Note that corporate firewall rule-sets are considered to be highly
sensitive, and were provided under non-disclosure agreements that
limit or delay our ability to discuss them publicly. Hence,
publishing quantified statistics about these rule-sets may be viewed a
contribution in itself.

An important contribution of this work is the identification and
selection of the~36 vendor-neutral firewall configuration errors to
search for. In principle, errors should be defined in terms of each
corporation's network security policy.  However, our stance is that of
an external auditor, and we do not really know these policies.
Therefore, the selected errors are all clear violations of well
established practices and guidelines (cf. \cite{sans-top20},
\cite{pci}, \cite{nist}).

The study's first finding is that serious errors are still alarmingly
frequent: For instance, the infamous Microsoft services, that are a
vector to numerous Internet worms (cf.\ \cite{CA2003-20}), are allowed
to {\em enter} organiztions' networks from the outside in 42\% of the
surveyed firewalls.  The most frequent error (allowing outbound SMTP
from over 256 IP addresses\footnote{This setting may be appropriate
  for small corporations, but is a significant error in larger
  corporations that run their own mail servers; Letting so many
  machines send e-mail directly, bypassing the corporate mail server, indicates
  that the corporation is not properly controlling its outbound
  e-mail.})  occured in over 80\% of the surveyed firewalls.
Moreover, since we only consider violations of ``best practices'' and
do not attempt to understand each network's semantics, the findings
should be viewed as a rough baseline: the protection offered by the
firewalls we surveyed may well be worse than suggested by this work.

One of the findings of \cite{woo04-conf} was that \CKPs\ with versions
4.1 or later had slightly fewer configuration errors than those with earlier
software versions. In contrast, the current study does {\em not\/}
show this effect---neither for \CKPs\ nor for \PIXs. The data shows that the
distribution of the number of configuration errors is essentially independent
of the firewall software version.  

Probably the most important finding of \cite{woo04-conf} was that
rule-set complexity was positively correlated with the number of
detected errors---and for this purpose \cite{woo04-conf} defined a
rule-set complexity measure denoted by \RC. Unfortunately, the \RC\ measure is
not suitable for \PIXs. Therefore, in order to
compare the complexity of configurations by two different vendors,
this work introduces a novel complexity measure, called the {\em
  firewall complexity} (\FC), that applies to both types of firewalls,
and should be suitable to other firewall vendors' configuration
languages as well.  A key feature of the \FC\ measure is that it
produces comparable values between \CKP\ and \PIX. Furthermore, the
\FC\ values are comparable to the older \RC\ values on \CKP.

What we shall see is that a rule-set's complexity, as measured by the
new \FC\ measure, is (still) positively correlated with the number of
detected configuration errors.  These findings hold for rule-sets from
both vendors. Thus we validate and strengthen the finding of
\cite{woo04-conf}, that, for well-configured firewalls, ``small is
(still) beautiful''. 

\eat{
However, unlike \cite{woo04-conf}, there is no
significant indication that later software versions have fewer errors
(for both vendors). This is apparently because the vendor-neutral
errors that this study focuses on are all controlled by the firewall's
basic filtering capability---which has not changed significantly
between versions.
}

The study also allows a preliminary investigation on whether the
choice of vendor had an impact on the quality of the configuration.
One finding is that \CKP\ configurations tend to be more complex than
\PIXs. Another finding is that in general \PIXs\ have
somewhat fewer errors than \CKPs\ of comparable \FC\ complexity.
However, the larger \PIXs\ are essentially as poorly configured as the
worst \CKPs. 

\noindent{\bf Organization:} 
The next section describes the data collection, provides some basic
statistics of the data, indicates some caveats in the methodology, and
describes the new \FC\ measure.  Section~\ref{sec:errors} describes
the selection criteria for the configuration errors used in this
study.  Section~\ref{sec:analysis} analyzes the data, describes the
detected error frequencies, discusses the effects of the firewall's
software version and rule-set complexity, and also compares the
relative complexity and number of errors between the two vendors.  We
conclude in Section~\ref{sec:conc}. Appendix~\ref{sec:list-of-errors} includes the full
list of possible configuration errors that were used.

\section{Methodology} \label{sec:methodology}

\subsection{\CKPs}

The data for this work includes 54 \CKP\ rule-sets that
were collected between 2003--2005. The rule-sets came from a variety
of organizations, from the telecommunications, financial, energy,
media, automotive, and healthcare market segments. 
Some basic statistics of the rule-sets are shown in
Table~\ref{tbl:ckpstats}.

\begin{table}[t]
  \begin{center}
    \begin{tabular}{|l|c|c|c|}
      \hline
          & Minimum & Median & Maximum   \\
      \hline
      \hline
        \#Rules    & 2 & 79 & 617 \\
      \hline
        \#Objects  & 19 & 572 & 5443  \\
      \hline
        \#Interfaces & 2 & 4 & 18 \\
      \hline
    \end{tabular}
  \end{center}
\caption {Statistical properties of the collected \CKP\ rule-sets. \#Rules
          counts the total number of rules in the rule-set (including
          NAT rules). \#Objects counts the number of network objects
          (hosts, subnets, and groups of these) defined in the
          database supporting the rules. \#Interfaces counts the
          number of network interface cards on the firewall.} 
 \label{tbl:ckpstats}
\end{table}

The collected rule-sets came from three Check Point major software
releases: 4.0, 4.1, and NG, with the NG rule-sets spanning several
minor releases. For this study I grouped all NG rule-sets up to
field-pack 3 in one category called ``NG/NG-FP3'', and all later
rule-sets in a separate category called ``NG R55''.  See
Table~\ref{tbl:ckpver} for a distribution of versions.  The firewall
version is relevant to our discussion since \cite{woo04-conf} reported
that rule-sets collected from later \CKP\ versions tended to have
fewer configuration errors---see Section~\ref{sec:fwver} for a
discussion.

\begin{table}[t]
  \begin{center}
    \begin{tabular}{|c|c|c|c|}
      \hline
   \multicolumn{4}{|c|}{\sc \CKP\ Version} \\
      \hline
          4.0 & 4.1 & NG/NG-FP3 & NG R55   \\
      \hline
          4 (7.4\%) & 30 (55.6\%) & 17 (31.5\%) & 3 (5.5\%) \\
      \hline
    \end{tabular}
  \end{center}
\caption {Distribution of \CKP\ rule-sets by software version. For
  each software version we show both the absolute number of rule-sets
  and their percentage (in parenthesis)}
 \label{tbl:ckpver}
\end{table}

\begin{table}[t]
  \begin{center}
    \begin{tabular}{|l|c|c|c|}
      \hline
          & Minimum & Median & Maximum   \\
      \hline
      \hline
        \#Lines  & 71 & 365 & 3259  \\
      \hline
        \#Interfaces & 2 & 4 & 8 \\
      \hline
    \end{tabular}
  \end{center}
\caption {Statistical properties of the collected \PIX\ rule-sets. \#Lines
          counts the total number of lines in the configuration file. 
          \#Interfaces counts the
          number of network interface cards on the firewall.} 
 \label{tbl:pixstats}
\end{table}

\begin{table}[t]
  \begin{center}
    \begin{tabular}{|c|c|c|c|}
      \hline
   \multicolumn{4}{|c|}{\sc \PIX\ Version} \\
      \hline
          4.4 & 5.0--5.2 & 6.0--6.2 & 6.3--7.0   \\
      \hline
          3 (10\%) & 7 (23.3\%) & 11 (36.7\%) & 9 (30\%) \\
      \hline
    \end{tabular}
  \end{center}
\caption {Distribution of \PIX\ rule-sets by software version. For
  each software version we show both the absolute number of rule-sets
  and their percentage (in parenthesis)}
 \label{tbl:pixver}
\end{table}

\subsection{\PIXs}

The data for this work also includes 30 \PIX\ rule-sets that
were collected between 2003--2005. 
Some basic statistics of the rule-sets are shown in
Table~\ref{tbl:pixstats}. 
The collected rule-sets 
include files from \PIX\ versions 4.4--7.0. Since there were
very few rule-sets from version 7.0 I grouped the rule-sets into four
categories: 4.4, 5.0--5.2, 6.0--6.2, and 6.3--7.0.
Table~\ref{tbl:pixver} shows the distribution of rule-sets into
versions. See Section~\ref{sec:fwver} for a discussion.

\subsection{Caveats}

When we consider the significance of the findings, we have to bear in
mind several caveats. 

The current study includes a total of 84 rule-sets. This is more than
double the sample of \cite{woo04-conf}, but still a very small
sample---The number of operational firewalls in the world is estimated
to be hundreds of thousands. Furthermore, these rule-sets were not
selected randomly: they were provided by organizations that wanted to
perform an audit of their firewall rule-set, and were willing to
purchase software for this purpose. Ancdotal evidence suggests that,
in some cases, these organizations felt their firewall rule-sets were too
complex to manage effectively without better software systems.
Therefore, our sample may be biased toward complex, badly configured,
firewalls.

In defense, I claim that the arguments of \cite{woo04-conf} are still
valid: The findings agree with the experience of many colleagues in
the network security industry, and the poor state of firewall
configuration can be observed, indirectly, by the huge proliferation
of network worms like Blaster~\cite{CA2003-20} and
Saphire~\cite{saph03} that could have been easily blocked by a well
configured firewall.

Finally, obtaining and reporting on {\em any} number of real
rule-sets from operational firewalls is extremely rare: These are
considered to be highly sensitive files. So there is significant
value in analyzing the data that {\em can} be obtained.

\subsection{A New Measure of Firewall Complexity}

One of the discoveries of \cite{woo04-conf} was that more complex
rule-sets have more errors in them. To make such statements precise,
in \cite{woo04-conf} I introduced a measure of complexity defined as
\[
  \mbox{\RC} = \mbox{\#Rules} + \mbox{\#Objects} +
  {\mbox{\#Interfaces}\choose 2},
\]
where \#Rules is the raw number of rules in the rule-set, \#Objects is
the number of network objects, and \#Interfaces is the number of
interfaces on the firewall. I believe that the \RC\ measure was quite
successful in capturing the essence of rule-set complexity. However,
for this study I realized that I needed to improve upon it, for two
reasons. First, I found that \RC\ was not suitable for \PIXs, and
second, I found that \RC\ didn't discriminate between more and less
complex rule-sets as well as I initially hoped.

\subsubsection{A comparable measure}
In order to compare the complexities of \PIXs\ to those of \CKPs\ I
needed a measure that produced comparable values for the two vendors.
However, I realized that \RC\ cannot be applied to \PIX, since
\PIX\ and \CKP\ are configured differently (see \cite{woo04-dir} 
for details). For the purpose of
measuring complexity, the most important differences are:
\begin{enumerate}
\item A \PIX\ configuration includes a separate set of rules
  (called an {\tt access-list}) for each
  {\em interface}, whereas a \CKP\ has a single set of rules that applies to all
  the interfaces---making \#Rules incomparable.
\eat{
\item A \PIX\ rule (i.e., an {\tt access-list} entry) is simpler than
  a \CKP\ rule: it has a single source object, a single destination
  object, and a single service, whereas a \CKP\ rule can have multiple
  objects in each category. Thus \PIXs\ tend to have many more rules
  than \CKPs---making the difficulty with \#Rules worse.
}
\item \PIX\ configurations do not have a separate object database, and
  up to version 6.0, \PIX\ configurations did not even allow the
  definition of non-trivial objects, such as objects containing
  anything other than a single subnet. Thus \#Objects is ill-defined
  for \PIXs. 
\end{enumerate}

\subsubsection{More interfaces imply higher complexity}
Beyond the immediate need for a measure that is suitable for \PIXs, I
realized that \RC\ could be improved upon for \CKPs\ as well.  
\eat{
The
main advantages of \RC\ were that it covered what I consider to be the
three most important parameters of complexity, and the relative values
it assigned to firewalls generally matched the intuition of firewall
administrators.  
}
Specifically, I felt that \RC\ didn't give enough weight to the number
of interfaces: as Table~\ref{tbl:ckpstats} shows, none of the surveyed
\CKPs\ has more than~18 interfaces (the median number was 4), yet it is
not uncommon to find \CKPs\ with hundreds of rules and thousands of
objects.  Since \#Interfaces is only {\em added\/} to the \RC, albeit
quadratically, its contribution is often dwarfed by the two other
terms. For instance, I felt that a firewall with 12 interfaces is {\em
  much\/} more complex than one with 3~interfaces with the same
numbers of rules and objects---and this intuition is not well captured
by the old \RC\ measure since the growth in the number of interfaces
only contributes an extra $(66 - 3) = 63$ points to the \RC\ of the
more complex firewall.

\subsubsection{The new \FC\ measure}
For all these reasons I decided to design a new measure of complexity.
To differentiate the new measure from the old I
call the new measure the {\em firewall complexity}, denoted \FC. Of
course, this new measure still had to be objective, intuitive, and
simple to compute.

For \PIXs, the simplest measure of complexity, and one which is
often used in discussions with \PIX\ administrators, is the number of
lines in the configuration file. This was my first attempt to pick a
new measure of complexity. However, I found the raw number of lines to
be slightly misleading, especially for very small configurations.
This is because even the smallest \PIX\ configuration file includes a
few tens of ``boilerplate'' lines that have little to do with traffic
filtering. To compensate for this, I chose to use the following definition. 
\begin{definition}
  Let \#Lines denote the number of lines in the ASCII file containing
  the complete \PIX\ configuration file. Then the {\em firewall
    complexity} of a \PIX\ firewall is:
\[
  \mbox{\FC}_p = \mbox{\#Lines} - 50.
\]
\end{definition}

To define a comparable measure for \CKPs, I tried to capture the
results of an imaginary over-simplified ``Check Point-to-PIX
converter''. Such a convertor would need to replicate the single \CKP\
rule-set and place a copy on each \PIX\ interface. Therefore, for the
\FC\ measure I chose to {\em multiply} the number of rules by the
number of interfaces. However, object definitions in (newer versions of)
\PIX\ are global (not per-interface), so \FC\ only needs to {\em
  add\/} add the number of \CKP\ objects, once. These choices have the
nice side-effect of giving much more weight to the number of
interfaces than the old \RC\ did, thereby addressing a shortcoming of
the \RC\ measure.  Thus, we arrive at the following definition.
\begin{definition}
  Let \#Rules denote the raw number of rules in the \CKP\ rule-set, let 
\#Objects denote the number of network objects, and let \#Interfaces
denote the number of interfaces on the firewall. Then the {\em firewall
  complexity} of a \CKP\ firewall is:
\[
  \mbox{\FC}_c = (\mbox{\#Rules}\times\mbox{\#Interfaces}) + \mbox{\#Objects}
\]
\end{definition}

\begin{figure}
 \begin{center}
  \resizebox{!}{3in}{\includegraphics{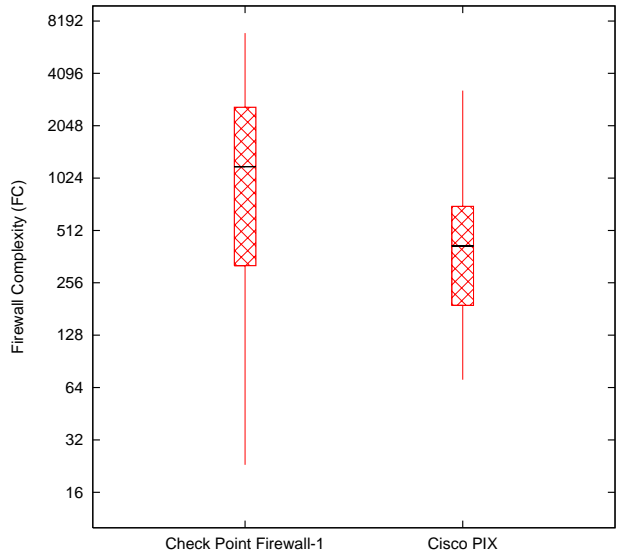}}
  \caption{Firewall complexity distribution (log scale) separately for
    \CKP\ and \PIX.  For each vendor we see a ``bar and whiskers''
    column for the distribution of errors. The bottom and top of the
    whiskers mark the minimum and maximum values, the bottom and top
    of the bar mark the first and 3rd quartiles, and the black line
    within the bar marks the median.  }
  \label {fig:complexity-brand}
 \end{center}
\end{figure}

Figure~\ref{fig:complexity-brand} shows the distribution of the
rule-set complexity as measured by \FC\ over the surveyed \CKP\ and
\PIX\ firewalls.  While the precise value
assigned by the new \FC\ to a \CKP\ rule-set differs from the value
computed by the old \RC\ for the same rule-set, the figure shows
that the genaral range of values remains the same: 10--10,000. But
more importatly, the \FC\ values assigned to \PIX\ rule-sets fall into
the same range as well.

The figure also shows that \CKP\ firewalls typically have a
higher complexity: the median \FC\ value for \CKPs\ is 1117 versus 315
for \PIX\ (note that the Y-axis is log-scaled). 
My interpretation is that this is a real finding rather than 
an artifact of the \FC\ metric: From inspecting the configuration data
directly, it seems to me that indeed \CKP\ configurations tend to be
more complex.

\section{The Selection of Configuration Errors} \label{sec:errors}

\subsection{Selection Goals and Criteria}

In \cite{woo04-conf} I considered twelve configuration errors. However,
eight of those errors were specific to Check Point firewalls,
which makes them unsuitable in a multi-vendor study. Therefore,
in this study I used a different list of errors. The current 
list is much larger list, consisting of 36 errors. All the errors are
vendor-neutral and create a risk to the network behind the firewall.
The full list of errors appears in Appendix~\ref{sec:list-of-errors}.

As in \cite{woo04-conf}, I took the stance of an external auditor.
Thus, the errors I counted are all violations of well established
practices and guidelines (cf. \cite{sans-top20},
\cite{pci}, \cite{nist}), that
are independent of each organization's specific requirements and
policies. As such, the findings should be viewed as a rough baseline:
The protection offered by the surveyed firewalls may well be worse
than suggested by this work.

Note that a single badly written rule may trigger multiple counted errors,
since some errors contain other, more specific, errors. For instance,
allowing Telnet (TCP on port 23) 
is a special case of allowing ``All TCP ports'', which
in turn is a special case of allowing ``Any service''. To avoid
inflating the error counts because of this containment phenomenon, a
more specific error was only counted if it was triggered by some rule
that does not trigger a more general error. Continuing the previous
example, a rule-set {\em can} trigger both the ``Any
service'' and the ``All TCP'' errors --- but only if the configuration
includes two separate rules, one allowing Any service, and the
other allowing all TCP ports.

One may argue that not all the configuration errors in the list are
equally severe: For instance, some services have a poorer security
history than others; also having all 65536 TCP ports open is probably
more risky than having just the Telnet port open. However, in order to
arrive at a single, easy to understand, number, I did not take such
fine-grain considerations into account. The simplicity of counting
each error type once allowed me to avoid introducing severity levels,
and let me side-step the question ``How many medium risk errors is one
high risk error worth?''.

Furthermore, one may argue that a configuration with ten rules
allowing Any service is riskier than a configuration with just one
such rule. My opinion is that the number of erroneous rules is a good
measure for the level of effort that the firewall administator would
need to spend {\em fixing\/} the configuration, but less indicative of
the amount of {\em risk\/}: If an attacker can enter the network, then
the network is at risk---and it is less important if the attacker has
one or ten firewall rules that allow him in. Therefore, again in the
interest of simplicity, I ignored the number of rules contributing to
each error, and opted for Boolean indicators for each error.

\begin{figure}[t]
 \begin{center}
  \resizebox{6in}{!}{\includegraphics{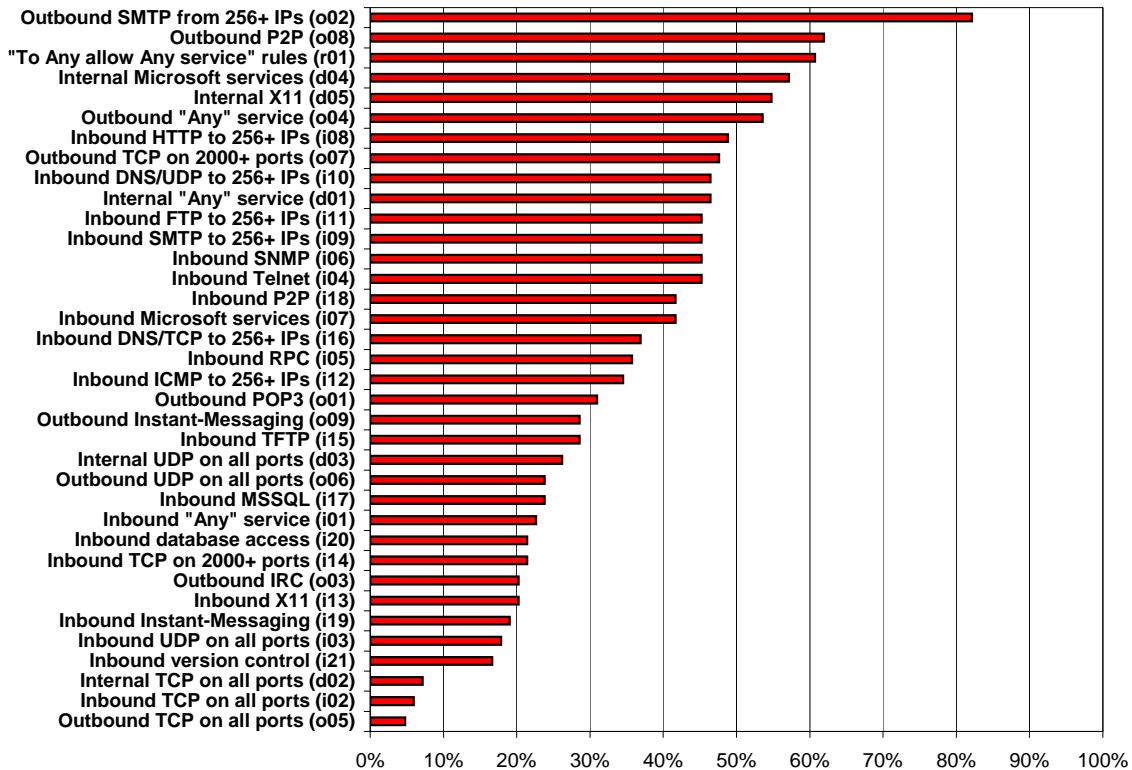}}
  \caption{Distribution of configuration errors.}
  \label {fig:findings}
 \end{center}
\end{figure}

\subsection{Error Categories}

The counted configuration errors (refer to
Appendix~\ref{sec:list-of-errors})
are organized into four categories:
inbound traffic, outbound traffic, internal traffic, and inherently
risky rules. The errors in each category have an identifying code with
a leading letter indicating the category: ``i'' errors for inbound
traffic, ``o'' errors for outbound traffic, ``d'' errors for internal,
and ``r'' errors for risky rules. 

The majority of configuration errors (21 of the~36) are for inbound
traffic (``i'' errors). This is in line with the firewall's perceived
primary job, which is to protect the internal resources from outsider
attach. These errors cover things such as allowing Any traffic
inbound, or allowing services such as NetBIOS, Telnet, RPC, SNMP, and
several other known-risky services, through the firewall in the
inbound direction.

I also wanted to count errors that are related to services such as
HTTP, DNS, and FTP. These are services that are often necessary to the
running of business, so allowing then inbound should not be counted as
an error {\em per se}. For instance, HTTP must be allowed to reach the
organization's web servers. However, allowing HTTP to reach machines that
are not hardened web servers is risky---many sensitive machines
(like e-mail servers, phone switches, routers, and firewalls)
provide a web interface.

The challenge is that we do not necessarily know which IP addresses
belong to the relevant servers and which do not. To get around this
challenge, I chose to use {\em thresholds}. The key idea is that it is
very rare to find hundreds of different web servers on one network.
Thus, if HTTP is allowed to reach over 256 IP addresses---more than a
full class-C subnet---this is almost always the result of badly
written rules. Several of the inbound traffic configuration errors
include such thresholds.

Configuration errors for outbound traffic (``o'' errors) contribute 
9~of the~36 items. Most of these deal with services that are
considered to be risky in all directions. In addition I counted errors
for indiscriminate outbound e-mail traffic (again using the notion of a
thresholds) and for IRC, which is notorious for carrying the
command-and-control channel of ``bot-nets''.

Internal traffic through the firewall (between separate internal
network segments) contributes~5 of the~36 items (``d'' errors): these
all deal with services that are considered to be risky in all
directions. Finally, I also counted a special ``r'' error for the
presence of rules that are inheretly problematic: An error is counted
if the rule-set has ``To Any allow Any service'' rules.

\section{Analysis and Discussion} \label{sec:analysis}

\begin{figure}
 \begin{center}
  \resizebox{!}{3in}{\includegraphics{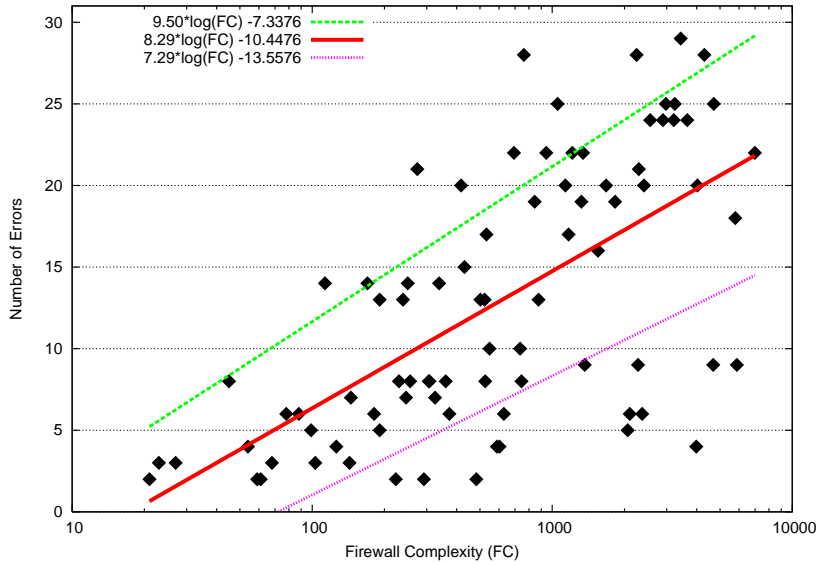}}
  \caption{Number of errors as a function of the rule-set's
    complexity \FC\ (log scale). The central red line represents the
    least-squares linear regression fit, and the green and purple 
    lines represent one
    standard deviation above and below the least-squares fit.}
  \label{fig:complexity}
 \end{center}
\end{figure}

\subsection{Firewalls are still badly configured}

Figure~\ref{fig:findings} shows the raw distribution of configuration
errors that were discovered in the data. The results are perhaps
unsurprising in view of~\cite{woo04-conf}: Generally speaking, the
surveyed firewalls are poorly configured. 

In the inbound direction, over 45\% of the firewalls allowed DNS, or FTP, or
SMTP, to reach over 256 addresses (items i10, i11, and i09
respectively). Possibly more worrisome is that 42\% of firewalls
allowed inbound NetBIOS traffic (item i07).

In the outbound direction the situation is apparently worse. Over 80\%
of firewalls allow broad outbound SMTP access (item o02)---and over
60\% of firewalls allow outbound P2P (o08), services which hardly ever
have any business use.  Finally, over 60\% of firewalls have rules of
the form ``from somwhere to Any allow Any service'' (r01): very lax
rules, which constitute gross mistakes by any account.

\subsection{Complexity Matters: Small is Still Beautiful} \label{sec:small}

\begin{figure}[t]
 \begin{center}
  \resizebox{4in}{!}{\includegraphics{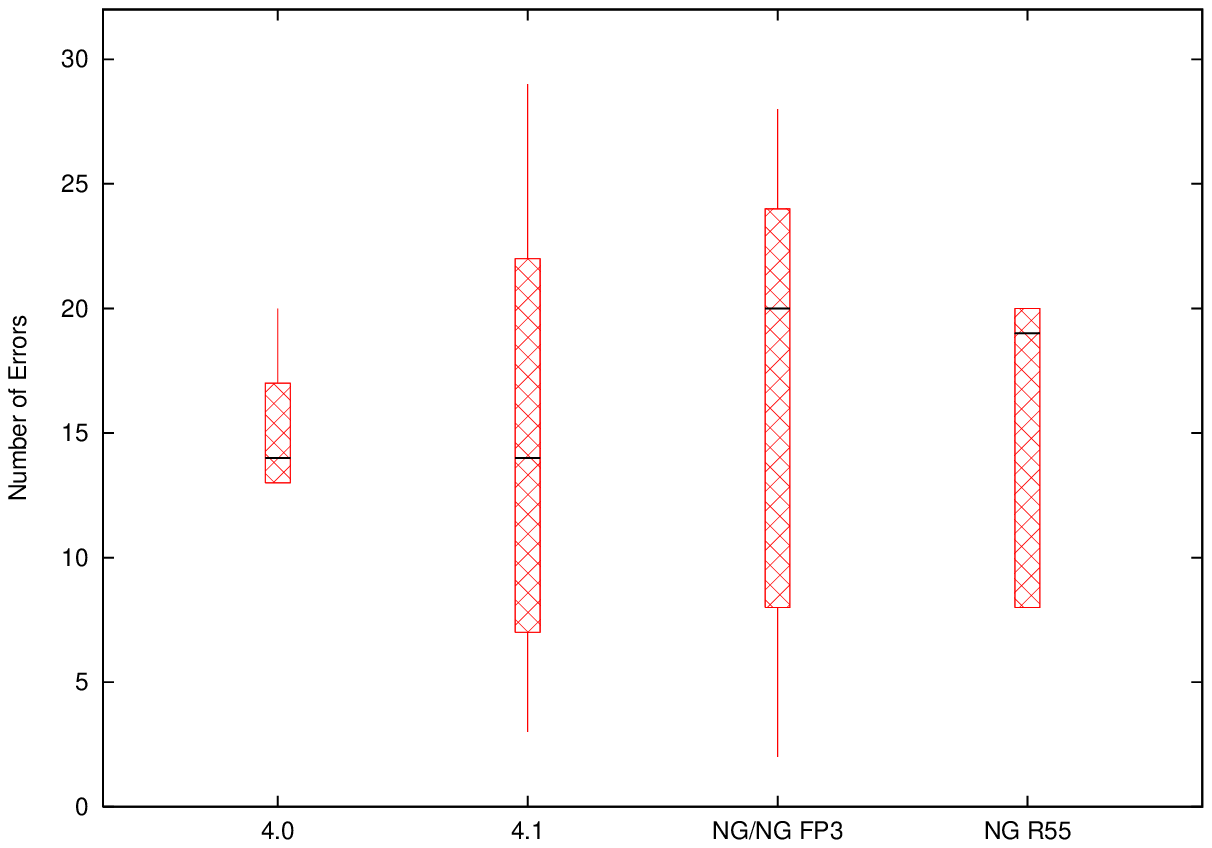}}
  \caption{Errors versus version --- comparing the distribution of
    errors between \CKP\ versions 4.0, 4.1, NG up to FP3, and NG R55. 
    For each software version
    we see a ``bar and whiskers'' column for the distribution of
    errors. The bottom and top of the whiskers mark the minimum and
    maximum values, the bottom and top of the bar mark the first and
    3rd quartiles, and the black line within the bar marks the median.
    }
  \label {fig:risk-ckp}
 \end{center}
\end{figure}

\begin{figure}
 \begin{center}
  \resizebox{4in}{!}{\includegraphics{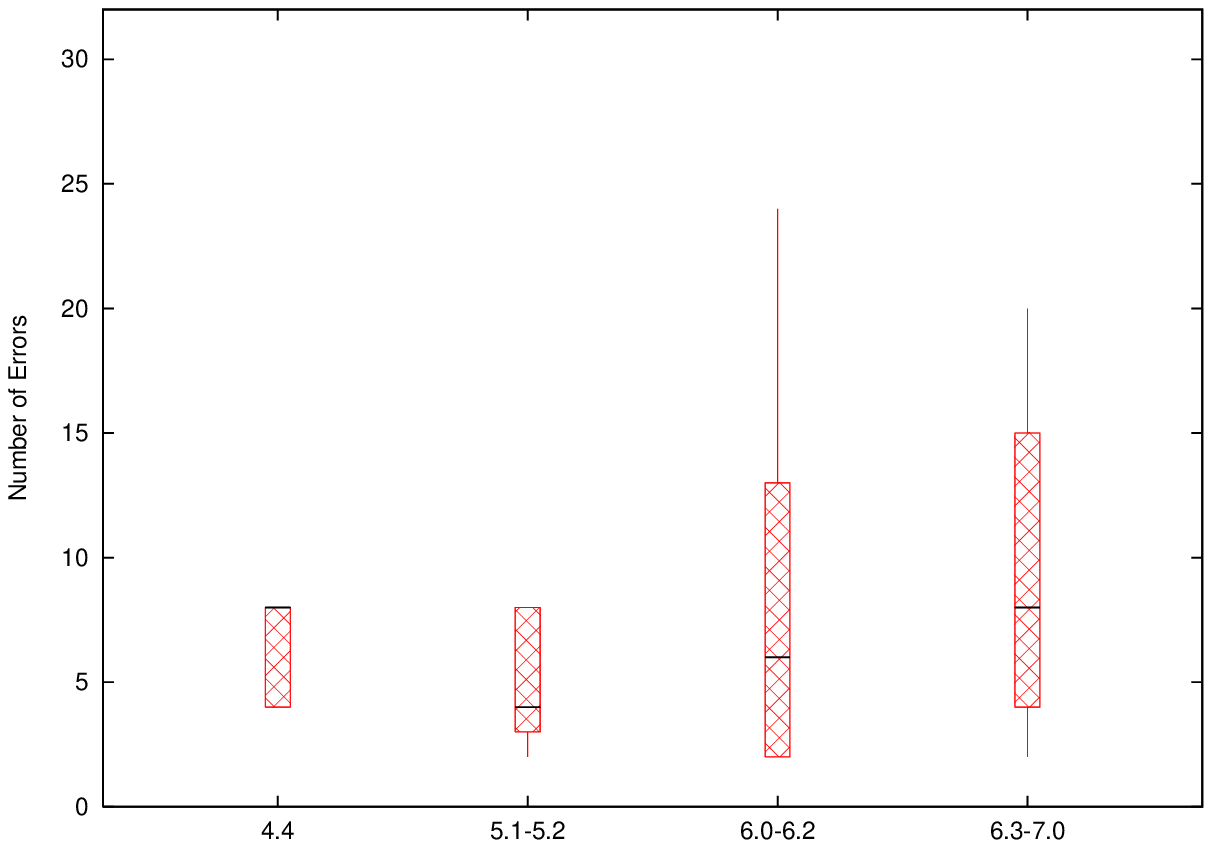}}
  \caption{Errors versus version ---
    comparing the distribution of
    errors between \PIX\ versions 4.4, 5.x, 6.0-6.2, and 6.3-7.0.
    For each software version
    we see a ``bar and whiskers'' column for the distribution of
    errors. The bottom and top of the whiskers mark the minimum and
    maximum values, the bottom and top of the bar mark the first and
    3rd quartiles, and the black line within the bar marks the median.
  }
  \label {fig:risk-pix}
 \end{center}
\end{figure}

One of the main findings of \cite{woo04-conf} was that high rule-set
complexity was positively correlated with the number of detected
configuration errors. Figure~\ref{fig:complexity} shows a scatter plot of the number
of errors versus \FC, and demonstates the same phenomenon in the
current survey. We can observe that there are very few high-complexity
rule-sets that are well-configured (the lower-right quadrant of the
figure is very sparse). Furthermore, the figure clearly shows a
correlation between the rule-set complexity, as measured by \FC, and
the number of detected errors.

In fact, the \FC\ measure gives us a crude but fairly accurate
prediction of the number of configuration errors: a linear regression
(the central line on Figure~\ref{fig:complexity}) shows that the
number of errors in a rule-set of complexity \FC\ is roughly captured
by the following formula:
\[\mbox{\#Errors} \approx 8\log_{10}(\mbox{\FC})-10.\] 

We can see that the current survey validates the finding of
\cite{woo04-conf}: it's safer to limit the complexity of a firewall
rule-set. Instead of connecting yet another subnet to the main
firewall, and adding more rules and more objects, it seems preferable
to install a new, dedicated, firewall to protect only that new subnet.
Complex firewall rule-sets are apparently too complex for their
administrators to manage effectively. 

As an aside, note that both Check Point and Cisco (and indeed other
major firewall vendors) are now offering virtualized solutions, in
which a single piece of hardware can be set up to function as multiple
separate firewalls. Besides offering various operational and commercial 
advantages, our results suggest that using multiple small (virtual) 
firewalls is likely to be more secure than a single large
firewall.

\subsection{Does the Firewall's Version Matter?} \label{sec:fwver}

One of the findings of \cite{woo04-conf} was that \CKPs\ with versions
4.1 or later had slightly fewer configuration errors than those with
earlier software versions.  I wanted to test whether the same trend
can be observed in this study as well.

Figures \ref{fig:risk-ckp} and \ref{fig:risk-pix} show the
distribution of errors across the different versions, separately for
\CKP\ and \PIX. It is clear from both figures that, contrary to the
findings of \cite{woo04-conf}, the effect of the software version on
the number of configuration errors is insignificant: the distribution
of the number of errors is essentially independent of the firewall
software version. The data does {\em not\/} support the hypothesis
that later software versions are correlated with fewer errors (for
both vendors).

I believe that \cite{woo04-conf} detected such a trend because 8~of
the configuration errors it considered were controlled by global options in the
\CKP\ user interface (rather than by explicit user-defined rules). The
default settings for those gloal options were improved in v4.1,
producing the detected effect.  However, all the configuration errors considered
in this study are vendor-neutral and are controlled by explicit
user-defined rules: i.e., by the basic filtering functionality of the
firewalls. This functionality has not changed syntactically or
semantically, in neither vendor's products, during the period that data was
collected. Therefore we see that later software versions do not
help users write better filtering rules.

\subsection{Does the Vendor Matter?} \label{sec:fwvendor}

\eat{
\begin{figure}
 \begin{center}
  \resizebox{3in}{!}{\includegraphics{errors-vs-brand.eps}}
  \caption{Errors versus Brand}
  \label {fig:risk-brand}
 \end{center}
\end{figure}

}

Given the fact that the data includes rule-sets from two vendors, with
rather different configuration languages and tools, I wanted to
investigate whether the choice of vendor had an impact on the quality
of the configuration. 

\begin{figure}
 \begin{center}
  \resizebox{!}{3in}{\includegraphics{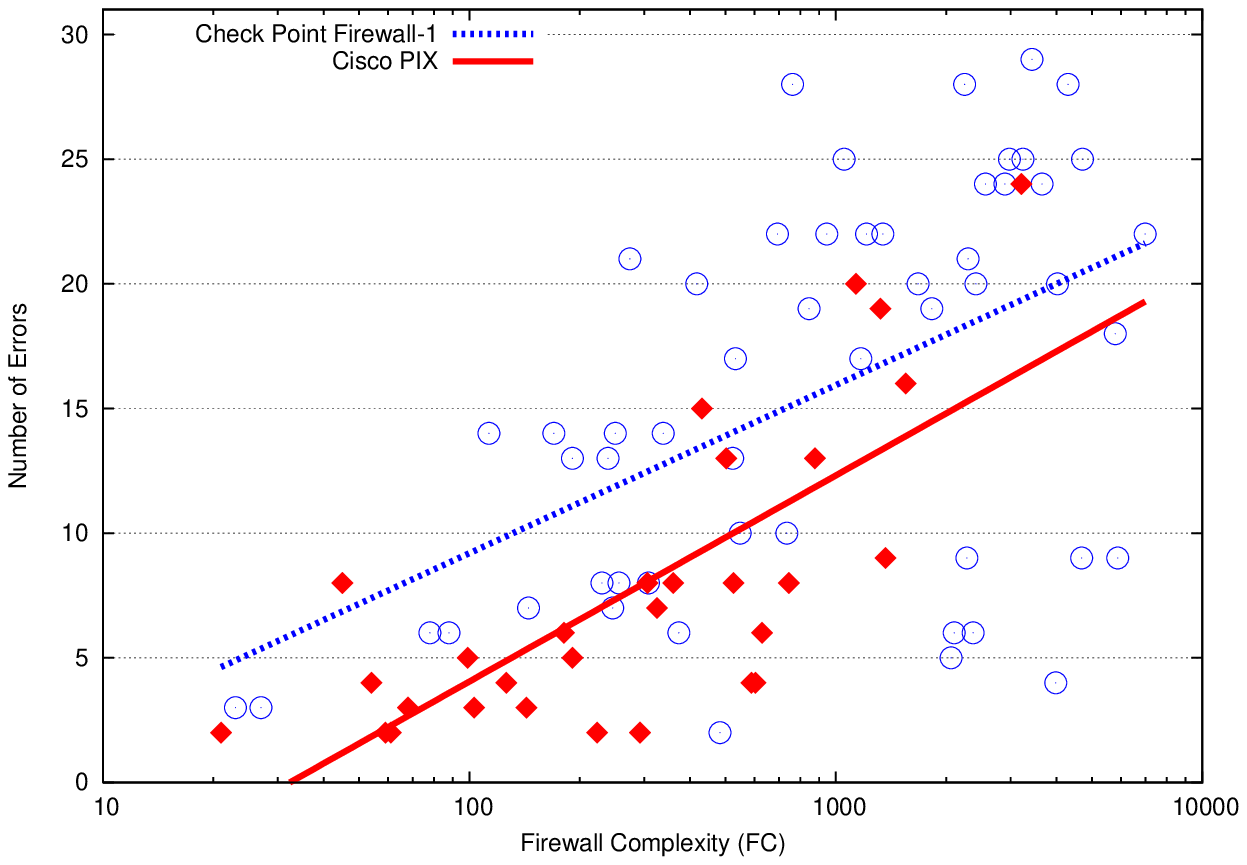}}
  \caption{Number of errors as a function of the rule-set's
    complexity \FC\ (log scale), separately for \CKP\ (blue circles) 
   and \PIX\ (red diamonds). The dotted blue line and solid red line
   represents the least-squares regression fit of the \CKPs\ and \PIXs, respectively.}
  \label {fig:separate-complexity}
 \end{center}
\end{figure}

\eat{
Figure~\ref{fig:risk-brand} shows the distribution of errors
separately for the two vendors. 
}

We already saw in Figure~\ref{fig:complexity-brand} that \CKP\
configurations tended to be more complex than \PIXs---and from
Figure~\ref{fig:complexity} we learned that more complex
configurations typically have more errors. Hence, we can expect 
\CKPs\ to have more errors than \PIXs\
just because of their greater complexity.  To
test the extent of this phenomenon, Figure~\ref{fig:separate-complexity} 
repeats the data of Figure~\ref{fig:complexity} but with separate
markers and regression lines for the two vendors.

As expected, the figure shows that in general \PIXs\ seem to exhibit
fewer errors than \CKPs: the median number of errors for the \PIXs\ is
6 versus~14 for \CKPs. Furthermore, the vertical distance between the
least-squares regression lines indicates that \PIXs\ have
approximately 5 errors less than \CKPs\ of comparable \FC\ complexity.
But a closer look reveals that small \PIXs, with \FC~$ < 1000$, indeed
have significantly fewer errors than small \CKPs. However, the larger
\PIXs, with \FC~$ \ge 1000$, are essentially as poorly configured as
the worst \CKPs. Note that this latter observation needs to be
validated with more data: the surveyed firewalls only include~5 \PIXs\
with \FC~$ \ge 1000$.

As background for a possible explanation, other sources of information
indicate that the \PIX\ configuration language and operational
environment is more difficult to learn and use than \CKP: E.g., \PIX\
involves a complex notion of ``direction'' (cf.\ \cite{woo04-dir});
\PIX\ mandates the use of network address translation\footnote{The
  strict requirement of writing NAT statements was relaxed with \PIX\
  version 7.x.} (NAT) statements---even
in the case that no translation is needed; Finally, many \PIXs\ are
managed from the command line whereas \CKPs\ are managed using a
graphical user interface. As a result, one often sees questions on how
to confiure the basic functionality of a \PIX\ on mailing lists such
as Firewall-Wizards \cite{fwwiz08}.

With this background, my interpretation of
Figure~\ref{fig:separate-complexity} is that, perhaps paradoxically,
the steeper learning curve and harsher work environment of a \PIX\
encourage administrators to produce smaller configurations---that are
less error-prone than comparable \CKPs. However, once the \PIX\
configuration complexity becomes large---the number of errors seems
similar to the numbers in comparable (complex) \CKPs.

\section{Conclusions} \label{sec:conc}

As we've seen, this work follows the methodology of \cite{woo04-conf}.
The current study is much larger, and is based on newer data, collected
from firewalls made by two major vendors, running later firewall
versions.  Finally, the study considers three times as many possible
vendor-neutral configuration errors. In order to compare the
complexity of configurations from different vendors, this work also
introduced a new uniform {\em firewall complexity} (\FC) measure, that
applies to both firewalls brands and probably to others as well.

Unlike \cite{woo04-conf}, the current study shows no significant indication
that later software versions have fewer errors (for either
vendor)---it seems that later software versions do not assist the
administrators to make fewer errors.

However, the findings do validate the main observations of
\cite{woo04-conf}: (a) firewalls are (still) poorly configured, and
(b) a rule-set's complexity is (still) positively correlated with the
number of detected configuration errors.  These findings hold for
rule-sets from both vendors. Thus we can conclude that, for
well-configured firewalls, ``small is (still) beautiful''.

{\small
\bibliographystyle {alpha}
\newcommand{\etalchar}[1]{$^{#1}$}

}

\appendix
\section{Appendix: The Selected Configuration Errors}  
  \label{sec:list-of-errors}

\subsection{Errors  Related to Inbound Traffic}

Twenty one of the errors I counted are related to traffic crossing the
firewall from the outside to the inside. These include:

\begin{description}

\item[i01: Inbound ``Any'' service] Allowing the ``Any'' service to enter
  the network is extremely risky since the ``Any'' service includes
  many vulnerable services.
\item[i02: Inbound TCP on all ports] Allowing TCP on all ports
  inbound is almost as risky as allowing Any service, since
  this includes many vulnerable TCP services.  On PIX firewalls, this
  risk is often the result of neglecting to specify a port number on
  an access-list or conduit statement: omitting the port number is
  interpreted as ``all ports''.
\item[i03: Inbound UDP on all ports] Similar to allowing all TCP ports
  but for UDP.
\item[i04: Inbound Telnet] Telnet is a remote-login service that is
  not encrypted. It is only authenticated by simple passwords, that
  are transmitted in the clear.  Anyone snooping the traffic can read
  all the transmitted information.
\item[i05: Inbound RPC] The RPC service is a mechanism for remote
  procedure calls in Unix environments. It is implemented using the
  {\tt portmapper} daemon, which assigns TCP ports to the services.
  RPC services have a long history of being insecure, and among others
  they include NFS (which potentially exposes all the organization's
  file system).
\item[i06: Inbound SNMP] SNMP is the Simple Network Management
  Protocol, which allows scanning and identifying the network
  infrastructure.  It is only authenticated by a simple password,
  called the "community string". The community string is often left at
  its default setting of "public", and transmitted in the clear.
  Several serious vulnerabilities have been reported in SNMP.
\item[i07: Inbound Microsoft services] NetBIOS is the name of a set of
  services that are used by the Microsoft Windows family of operating
  systems to support network functions such as file and printer
  sharing. These services are very insecure, and include some of the
  most attacked services \cite{sans-top20}. I counted an error if
  the firewall allowed any of these services to cross it.
\item[i08: Inbound HTTP to 256+ IPs] HTTP is the main protocol used
  for web browsing, so allowing it inbound is not counted as an error
  {\em per se}. However, allowing HTTP reach machines that are not
  hardened web servers is risky: Many sensitive machines provide a web
  interface: These include e-mail servers, printers, phone switches,
  routers, and firewalls. Therefore allowing HTTP to reach over 256 IP
  addresses is counted as an error because it is very rare to find so
  many web servers on one network---such access is almost always the result of
  badly written rules.
\item[i09: Inbound SMTP to 256+ IPs] SMTP is the main E-mail
  protocol---and E-mail is a vector for many viruses and worms.
  Nevertheless, allowing SMTP inbound is a requirement for many
  organizations, so inbound SMTP is not counted as an error. However,
  it is very rare to find more than 256 mail servers on one network,
  so allowing SMTP to reach over 256 IP addresses is counted as an
  error.
\item[i10: Inbound DNS/UDP to 256+ IPs] DNS is the Domain Name Service
  on UDP port 53 or on TCP port 53. DNS is one of the most attacked
  services in use \cite{sans-top20}. As before, inbound DNS is often
  required---so it's not counted as a risk---but allowing DNS to reach over
  256 DNS servers on one network {\em is}.
\item[i11: Inbound FTP to 256+ IPs] FTP is the File Transfer Protocol,
  which is necessary for many organizations. Serious vulnerabilities
  have been found in many versions of FTP server software. As before,
  an error is counted if FTP is allowed to reach over
  256 servers on one network.
\item[i12: Inbound ICMP to 256+ IPs] ICMP is the Internet Control
  Message Protocol, which is necessary for many network management
  tools (like {\tt ping}). Full ICMP access allows outsiders to scan
  the network, it lets worm-generated traffic cross the perimeter, and
  it can be used to mount denial-of-service attacks. An error is
  counted if ICMP can reach over 256 internal machines.
\item[i13: Inbound X11] X11 is a popular graphical window system
  developed at MIT and implemented on UNIX systems to allow UNIX-based
  applications to be run from multiple types of terminals, PCs, and
  workstations. X11 is not encrypted and uses weak authentication.  An
  error is counted if X11 is allowed inbound.
\item[i14: Inbound TCP on 2000+ ports] Allowing traffic over 2000 TCP ports
  inbound is almost as risky as allowing all TCP ports (item i02), since
  this includes many vulnerable TCP services. Such access is usually
  the result of an ``permit all TCP ports'' rule that is slightly
  mitigated by a few earlier ``deny TCP on port X'' rules. 
\item[i15: Inbound TFTP] TFTP is the Trivial File Transfer Protocol.
  It is not encrypted and not authenticated in any way.  Anyone who
  can use TFTP inbound can upload any file onto any device that
  responds to TFTP, such as routers, firewalls, and other
  communication equipment.
\item[i16: Inbound DNS/TCP to 256+ IPs] This is the same as item i10
  except over TCP.
\item[i17: Inbound MSSQL] The Microsoft SQL Server (MSSQL) contains several serious
vulnerabilities that allow remote attackers to obtain sensitive
information, alter database content, compromise SQL servers, and, in
some configurations, compromise server hosts. It's in the  
SANS Top 20 list \cite{sans-top20}.
\item[i18: Inbound P2P] Peer-to-peer services (napster, edonkey,
  gnutella, kazaa) are used to download, and distribute many types of
  data (e.g., music, video, graphics, text, source code, and
  proprietary information to name a few). These services are in the
  SANS Top 20 list \cite{sans-top20}.
\item[i19: Inbound Instant-Messaging] Instant Messaging services (MSN, yahoo, aol)
let users use text-based chat, check remote web based
email, do voice chat, perform video communication, and send and share
data files. These services are in the SANS Top 20 list \cite{sans-top20}.
\item[i20: Inbound database access] Database protocols (SQLnet, mysql,
  postgresql) provide remote access to relational databases. Allowing
  direct inbound access to your corporate databases is risky.  These
  services are in the SANS Top 20 list \cite{sans-top20}.
\item[i21: Inbound version control] Version control systems (cvs,
  subversion) provide tools to manage different versions of documents
  or source code. Both systems have serious known vulnerabilities.
  These services are in the SANS Top 20 list \cite{sans-top20}.

\end{description}

\subsection{Errors  Related to Outbound Traffic}

For outbound traffic I counted three specific services that are
particularly problematic in the outbound direction (o01, o02, and
o03), plus six errors that are similar to inbound
errors, as follows:

\begin{description}
\item[o01: Outbound POP3] POP3 is used to download E-mail from mail
  servers to desktop computers.  Since E-mail is a vector for viruses
  and worms, many organization only allow POP3 to reach their internal
  E-mail servers, and forbid any access to external E-mail servers. An
  error is counted if POP3 is allowed outbound.
\item[o02: Outbound SMTP from 256+ IPs] Outbound SMTP (E-mail) should only
  originate from mail servers, not from individual desktops. An error
  is counted if over 256 IP addresses can send outbound SMTP through
  the firewall.
\item[o03: Outbound IRC] Internet Relay Chat (IRC) is an interactive
  chat service. Many worms and Trojans use outbound IRC as a communications
  vector.
\item[o04: Outbound ``Any'' service] Similar to the i01 item.
\item[o05: Outbound TCP on all ports] Similar to the i02 item.
\item[o06: Outbound UDP on all ports] Similar to the i03 item. 
\item[o07: Outbound TCP on 2000+ ports] Similar to the i14 item.
\item[o08: Outbound P2P] Similar to the i14 item.
\item[o09: Outbound Instant-Messaging] Similar to the i19 item.
\end{description}

\subsection{Errors  Related to Internal Traffic}

For traffic between different internal network segments that are
separated by the firewall the errors are a small subset of the inbound
errors, as follows:

\begin{description}
\item[d01: Internal ``Any'' service]  Similar as the i01 item.
\item[d02: Internal TCP on all ports]  Similar as the i02 item.
\item[d03: Internal UDP on all ports] Similar as the i03 item.
\item[d04: Internal Microsoft services] Similar as the i07 item.
\item[d05: Internal X11] Similar as the i13 item.
\end{description}

\subsection{Inherently Risky Rules}

In addition to all the traffic-based errors described above, I also
counted a special error for rules that are inheretly problematic, even
if there are some other rules that mitigate some of their bad
effects. An error is counted (with code {\bf r01}) if the rule-set
has ``To Any allow Any service'' rules.

\end{document}